\documentclass[aps,prd,preprint,superscriptaddress]{revtex4-1}
\pdfoutput=1
\usepackage{graphicx}  
\usepackage{subfigure}
\usepackage{latexsym}
\usepackage{slashed}
\usepackage{dcolumn}   
\usepackage{bm}        
\usepackage{amsmath}
\usepackage{amssymb}   
\usepackage{longtable}
\usepackage[section]{placeins}
\usepackage{float}

\def\beq{\begin{equation}}
\def\eeq{\end{equation}}
\def\bey{\begin{eqnarray}}
\def\eey{\end{eqnarray}}
\newcommand{\be}{\begin{equation}}
\newcommand{\ee}{\end{equation}}
\newcommand{\bea}{\begin{eqnarray}}
\newcommand{\eea}{\end{eqnarray}}
\newcommand{\bma}{\begin{matrix}}
\newcommand{\ema}{\end{matrix}}

\newcommand{\bml}{\begin{mathletters}}
\newcommand{\eml}{\end{mathletters}}
\newcommand{\bes}{\begin{subequations}}
\newcommand{\ees}{\end{subequations}}

\newcommand{\bi}{\begin{itemize}}
\newcommand{\ei}{\end{itemize}}

\newcommand{\uva}{\affiliation{Department of Physics, University of Virginia, Charlottesville, VA 22904-4714, USA}}
\newcommand{\hue}{\affiliation{Center for Theoretical and Computational Physics, Hue University College of Education, Hue, Vietnam}}

\begin{document}





\title{Astrophysical Constraints on Inflationary Dark Matter in the Luminogenesis Model}

\author{Pham Q. Hung}
\email{pqh@virginia.edu}\uva \hue
\author{Kevin J. Ludwick}
\email{kludwick@virginia.edu} \uva






\begin{abstract}

The assumption of collisionless cold dark matter on its own cannot reconcile several astrophysical discrepancies (cusp-vs-core problem, 
missing satellite problem, too-big-to-fail problem).  Self-interacting dark matter provides a promising framework for solving all these problems, and self-interaction 
cross sections are duly constrained in the literature.  Following the 
work of Tulin, Yu, and Zurek \cite{Zurek}, we can constrain the dark matter mass and the mass of a light mediator assuming a generic scalar Yukawa-type interaction.  In particular, 
we constrain the strongly coupled inflationary dark matter of the luminogenesis model, a unification model with the gauge group $SU(3)_C \times SU(6) \times U(1)_Y$, 
which breaks to the Standard Model with an extra gauge group for dark matter 
when the inflaton rolls into its true vacuum.  The luminogenesis model is additionally subject to constraints on inflation, and we find an upper bound on the scale 
of symmetry breaking of the inflaton and the decoupling scale $M_1$ of certain representations of the gauge group.  
We emphasize that the luminogenesis model enables a unique connection between astrophysical 
constraints, the nature of dark matter, and inflation.  
\end{abstract}
\maketitle

\renewcommand{\thepage}{\arabic{page}}
\setcounter{page}{1}
\renewcommand{\thefootnote}{\#\arabic{footnote}}


\begin{center}
{\bf Introduction}
\end{center}

The formation of galaxies and galaxy clusters is heavily influenced by the nature of dark matter.  For the usual framework of cold dark matter, there are 
discrepancies between their predictions and 
observations.  $N$-body simulations for exclusive collisionless cold dark matter predict the central density profile of dwarf galaxy and galaxy cluster halos to be very cusp-like, whereas observations indicate flat cores (cusp-vs-core problem) \cite{Blok}.  The number of Milky Way satellites predicted in simulations is bigger by an order of 
magnitude than the number inferred from observations (missing satellite problem) \cite{Moore, Klypin}, although this may not be very troublesome if more ultra-faint galaxies are successfully detected in the future \cite{Simon}.  The brightest observed dwarf spheroidal galaxy satellites of the Milky Way are predicted to be 
in the largest Milky Way subhalos, but the largest subhalos are too massive to host them (too-big-to-fail problem) \cite{Kolchin}.  The resolution of these problems may come through 
several possible means, including more 
accurate consideration of baryon interactions, astrophysical uncertainties, and warm dark matter.  A promising framework that can solve all these issues is self-interacting 
dark matter. 

In this work, we are interested in astrophysical constraints on strongly coupled scalar Yukawa-type interactions, and our methodology follows that of \cite{Zurek}.  (And we 
explain our use of the term "Yukawa-type" later in the text.)  The astrophysical 
constraints on self-interaction cross sections confer restrictions on the allowed masses of dark matter and the scalar light mediator in the Yukawa-type interaction.  
When applied to inflationary dark matter of the luminogenesis
 model \cite{Luminogenesis1, Luminogenesis2, pqkevin}, these restrictions imply constraints on the unification scale of the model and the parameters of the inflation potential.  
 
 We first briefly review the luminogenesis model and its significance.  We then present our results constraining the dark matter and scalar mediator masses that come from 
 comparing the Yukawa-type interaction cross section from solving Schr\"{o}dinger's equation and astrophysical constraints.  Applying the results to the luminogenesis model, 
 we arrive at a unique, interesting connection between astrophysical bodies, dark matter, and cosmic inflation.  
 
 \bigskip
 \begin{center}
 {\bf Review of the Luminogenesis Model}
 \end{center}

In the luminogenesis 
model, the dark and luminous sectors are unified above the Dark Unified Theory (DUT) scale.  
At this DUT scale, the unified symmetry of the model breaks ($SU(3)_C \times SU(6) \times U(1)_Y \rightarrow SU(3)_C \times SU(4)_{DM} \times SU(2)_L \times U(1)_Y  \times U(1)_{DM}$), and the breaking is triggered by the inflaton's slipping into the minimum of its symmetry-breaking (Coleman-Weinberg) potential and acquiring the true vacuum expectation value $\mu_{DUT}$, 
which is the DUT scale energy.  This symmetry breaking allows the 
inflaton to decay to dark matter, and dark matter can in turn decay to Standard Model (SM) and "mirror" matter.  
The representations and group structure of the luminogenesis model for each of the three families are given below \cite{Luminogenesis2}.  The existence of 
"mirror" fermions, as proposed by \cite{PQmirror1}, is necessary for anomaly cancellation, and it provides a mechanism in which right-handed neutrinos may 
obtain Majorana masses proportional to the electroweak scale, and they 
could be searched for at the Large Hadron Collider.  
\begin{table}[h]
 \begin{center}
 \begin{tabular}{|l|l|lr|||} \hline
 $SU(6)$ & $SU(4)_{DM} \times SU(2)_L \times U(1)_{DM}$ \\ \hline
 ${\bf 6}$ & ${\bf (1,2)_2 + (4,1)_{-1}}$  \\
 $ {\bf 20}$ & ${\bf (4,1)_3 + (4^\ast , 1)_{-3} + (6,2)_{0} }$ \\
 ${\bf 35}$ & ${\bf (1,1)_0 + (15,1)_0 + (1,3)_0 +(4,2)_{-3}}$  \\ 
 & ${\bf  + (4^\ast , 2)_3}$ \\  \hline
 \end{tabular}
 \end{center}
 \caption{\label{table1} ${\bf (1,2)_2}$ represents luminous matter while ${\bf (4,1)_3 + (4^\ast , 1)_{-3}}$ represent dark matter.}
 \end{table}
 \begin{table}[h]
 \begin{center}
 \begin{tabular}{|l|l|lr|||} \hline
  & $SU(3)_c \times SU(6) \times U(1)_{Y}$ \\ \hline
 R $\supset$ SM fermions & ${\bf (3,6, 1/6)_L + (1,6, -1/2)_L }$  \\
 &${\bf + (3,1,2/3)_R + (3,1,-1/3)_R }$ \\ 
 &${\bf + (1,1,-1)_R}$ \\
 \hline
 R $\supset$ "mirror fermions" &  ${\bf (3,6, 1/6)_R + (1,6, -1/2)_R }$  \\
 &${\bf + (3,1,2/3)_L + (3,1,-1/3)_L }$ \\ 
 &${\bf + (1,1,-1)_L}$ \\
 \hline
 R $\supset$ dark-matter fermions & ${\bf (1,20,0)}$ \\  \hline
 \end{tabular}
 \end{center}
 \caption{\label{table2} R in the left column denotes representation. Standard Model (SM) left-handed doublets and right-handed singlets comprise the first row, 
 mirror right-handed doublets \cite{PQmirror1, PQmirror2} and left-handed singlets comprise the second row, and dark-matter left- and right-handed fermions belong to the last row.}
 \end{table}

Mirror quarks and leptons can be searched for at the LHC \cite{UVAOKS}. The search for the electroweak-scale right-handed neutrinos is particularly interesting since it will be a direct test of the seesaw mechanism. As emphasized in \cite{PQmirror1}, the production of $\nu_R \nu_R$ at the LHC can give rise to interesting signals such as like-sign dileptons. Furthermore, the lightest mirror quark can decay into a Standard Model (SM) quark by emitting a SM-singlet Higgs scalar ($q^M_R \rightarrow q_L + \phi_S$) through an interaction Lagrangian of the form $g_{Sq} \,\bar{q}_L \ \phi_S \ q_R^M + h.c.$, where $q_L$ and $q_R^M$ refer to a SM left-handed and mirror right-handed quark doublet respectively.  A similar decay process applies to the lightest mirror lepton ($l^M_R \rightarrow l_L + \phi_S$) with $g_{Sl} \,\bar{l}_L \ \phi_S \ l_R^M + h.c.$

The inflaton $\phi_{inf}$ is represented by ${\bf (1,1)_0}$ of $\bf{35}$, and since ${\bf 20 \times 20 = 1_s+35_a+175_s+189_a}$, the inflaton decays mainly into dark matter through the interaction $g_{20} \, \Psi_{20}^{T} \sigma_2 \Psi_{20} \, \phi_{35}$ which contains the inflaton in $g_{20} \, \chi_{L}^{T} \sigma_2 \chi^{c}_{L} \phi_{inf}$. This is one of the main points of \cite{Luminogenesis2}: the predominance of dark over luminous matter.  Dark matter can then be converted via the process of what \cite{Luminogenesis2} refers to as "luminogenesis."  We summarize this process below:
\begin{itemize}
\item As is noted in \cite{Luminogenesis2}, for the $SU(4)_{DM}$ dark matter (DM) fermion $\chi$, a small asymmetry in the number density $\Delta n_\chi=n_\chi - n_{\bar{\chi}}$ is assumed to be present, with $n_\chi = n_{sym} + \Delta n_\chi$ only slightly bigger than $n_{\bar{\chi}}= n_{sym}$, where $n_{sym}$ is the symmetric part of $n_{\chi}$ and $n_{\bar{\chi}}$.  The asymmetric part consists of the small excess $\Delta n_\chi \ll n_{sym}$.  For the origin of this asymmetry in the DM number density, we assume that there is a global $U(1)_{\chi}$ symmetry for DM. The interactions involving the gauge bosons of the coset group $SU(6)/SU(4) \times SU(2) \times U(1)_{DM}$ explicitly break the $U(1)_{\chi}$ symmetry, and their decays involving the interferences between the tree-level and one-loop diagrams will ultimately produce a net DM asymmetry, assuming the presence of CP violation in the DM sector.  (This process is similar to the one involving X and Y gauge bosons in $SU(5)$ Grand Unified Theory.)
\item $\chi$ and $\bar{\chi}$ can annihilate via $\gamma_{DM}$, the massive dark photon of $U(1)_{DM}$, into luminous particle-antiparticle pairs via the effective interaction $\frac{g^2}{M_{\gamma_{DM}}}(\bar{\chi}  \gamma_{\mu} \chi)(\bar{f} \gamma^{\mu} f)$, and the particle-antiparticle pairs of luminous fermions annihilate to radiation.  
\item $\chi$ and $\bar{\chi}$ can also be converted to luminous leptons via the interactions with two scalar fields: $\Phi_{15}^{(L)}$ and 
$\Phi_{\bar{15}}^{(R)}$, and given by $\frac{g_{6}^2}{M_{15}^2} \, (\chi^{T}_{L} \sigma_2 l_L)\, (\chi^{c,T}_{L} \sigma_2 l^{M,c}_{L}) + h.c.$, resulting in $\chi_L + \chi_R \rightarrow l_L + l^{M}_R $ and $\bar{\chi}_L + \bar{\chi}_R \rightarrow \bar{l}_L + \bar{l}^{M}_R $. Although the decay length of the previously-mentioned process $l^M_R \rightarrow l_L + \phi_S$ could be macroscopic at the LHC ("long-lived" $l^M_R$), in the early universe, $l^M_R$ basically decays promptly into SM leptons.
\item The coefficients for the annihilation process involving $\gamma_{DM}$ and the conversion process involving the scalars $\Phi_{15}^{(L)}$ and $\Phi_{\bar{15}}^{(R)}$ are independent of each other, and they are such that 14\% of all dark matter (14\% of asymmetric and symmetric parts) converts via the scalars and 86\% of asymmetric and symmetric parts annihilates via $\gamma_{DM}$ ultimately to radiation.  The 14\% of the symmetric parts of $n_\chi$ and $n_{\bar{\chi}}$ that is converted to luminous matter via the scalars has equal parts of luminous particles and anti-particles, and these annihilate to radiation.  So the whole of the symmetric parts of $n_\chi$ and $n_{\bar{\chi}}$ is mainly converted to radiation.  But since annihilation via $\gamma_{DM}$ requires the presence of both $\chi$ and $\bar{\chi}$, this annihilation does not affect the asymmetric part $\Delta n_\chi$.  So overall, we are left with radiation from the symmetric parts of $n_\chi$ and $n_{\bar{\chi}}$, and of the asymmetric part, 14\% is luminous matter and  86\% is DM, giving the correct proportion of luminous to dark matter as the conversion process via the scalars freezes out.
\end{itemize}

So we see that this process of luminogenesis depends upon both freeze-out and asymmetry in dark matter (and the asymmetry is propagated through to luminous matter), and the coefficients of the processes discussed are such that luminogenesis gives what is observationally expected.  These processes involved in luminogenesis are discussed in more detail in \cite{Luminogenesis2} and are being explored further in \cite{hunglumino}.

It is assumed that ${\bf (15,1)_0 + (1,3)_0}$ ${\bf +(4,2)_{-3}+(4^{\ast},2)_3}$ of ${\bf 35}$ and ${\bf (6,2)_0}$ of ${\bf 20}$ 
have masses that are on the order of the DUT scale and thus do not affect the particle theory below that energy scale.  
The $SU(4)_{DM}$ dark matter fermions are represented by ${\bf (4,1)_3 + (4^\ast , 1)_{-3}}$ in the ${\bf 20}$ representation of $SU(6)$.  
Since dark matter should have no $U(1)_Y$ charge, 
the $SU(4)_{DM}$ particles in ${\bf (4,1)_{-1}}$ in the ${\bf 6}$ representation of $SU(6)$ cannot be dark matter since they have $U(1)_Y$ charge (as shown in Table \ref{table2}), and they are assumed to decouple below the mass scale we 
call $M_1$.  In \cite{pqkevin}, we make predictions for the mass of $\chi$ in the following way:
\begin{itemize}
\item We run the $SU(2)_L$ gauge coupling from the known electroweak scale up to some unknown DUT scale where it intersects with the $SU(4)_{DM}$ gauge coupling $\alpha_4$.
\item Then we run $\alpha_4$ down to its confinement scale, which is when $\alpha_4 \sim 1$.  In analogy with QCD confinement of $SU(3)_C$, the main contribution to $SU(4)_{DM}$ fermions' dynamical mass is from the confinement scale of $SU(4)_{DM}$, and that energy scale is our mass prediction for $\chi$.  
\item In order to specify that scale, we need to specify a DUT scale.  Since $SU(6)$ breaks at the DUT scale when the inflaton slips into its true vacuum, we specify the DUT scale and therefore the dynamical mass of $\chi$ by constraining the parameters of a symmetry-breaking (Coleman-Weinberg) inflaton potential with Planck's constraints on the scalar spectral index and amplitude \cite{Planck, Plancki}.  
\end{itemize}
We will compare these predictions from \cite{pqkevin} to the results 
from this paper's application of astrophysical constraints to dark matter mass.

Because of the confinement of $SU(4)$, dark baryons are formed from four $\chi$ particles.  
These particles are dubbed CHIMPs, which stands for "$\chi$ Massive Particles."  A CHIMP is denoted by $X$, and $X=(\chi \chi \chi \chi)$.  As we know from QCD, $SU(3)$ Nambu-Goldstone (NG) bosons appearing from the spontaneous breaking 
of chiral symmetry from $<\bar{q} q> \neq 0$ acquire a small mass from the explicit breaking of 
quark chiral symmetry due to the small masses of quarks, and they become pseudo-NG bosons known as pions.  We expect a similar phenomenon from $<\bar{\chi} \chi> \neq 0$ in 
$SU(4)$, and the NG bosons can acquire a small mass through a term $m_{0} \bar{\chi} \chi$ with $m_{0}$ a parameter that is related to $m_{\pi_{DM}}$ and should 
be on the order of  $m_{\pi_{DM}}$, 
and $m_0 \ll \Lambda_4 \sim m_\chi$.  We seek to constrain the $m_{\pi_{DM}}$-$m_X$ parameter space through astrophysical constraints via the procedure in the following section.

\bigskip

\begin{center}
{\bf Solving Schr\"{o}dinger's Equation}
\end{center}

For unspecified $X$ and $\pi_{DM}$, in general, the cross section of their interaction may not lie in the regimes of the Born or classical approximations, so we cannot rely solely on analytical expressions for these regimes.  In order to find how the mass of strongly coupled DM is correlated to the mass of a scalar mediator via astrophysical constraints, we need to numerically solve Schr\"{o}dinger's equation, 
and we use the methodology described in detail in \cite{Zurek}.  

We take the interaction between DM (a CHIMP, denoted by $X=(\chi \chi \chi \chi)$) and a scalar mediator ($\pi_{DM}$) to be given by an attractive Yukawa-type potential
\beq
V(r) = - \frac{\alpha_{DM}}{r} e^{-m_{\pi_{DM}} r} \, , \label{potential}
\eeq
with interaction
\beq
\mathcal{L}_{\rm int} = g_{DM} \bar X X \pi_{DM},
\eeq
where $\alpha_{DM} = g_{DM}^2/(4\pi)$.  We use the term "Yukawa-type" because a Yukawa interaction is typically thought of as between a spin-$\frac{1}{2}$ particle and a scalar, whereas 
we are using Eq. (\ref{potential}) to describe the interaction between $X$, a boson, and the scalar $\pi_{DM}$.  This distinction does not make a difference in calculation in our 
case since Schr\"{o}dinger's equation does not account for spin.  
In analogy with QCD's strong nucleon-pion interaction, we assume that $\alpha_{DM} \sim 1$. 

\begin{figure}[h]
\begin{center}
\fbox{\includegraphics[scale=1.4]{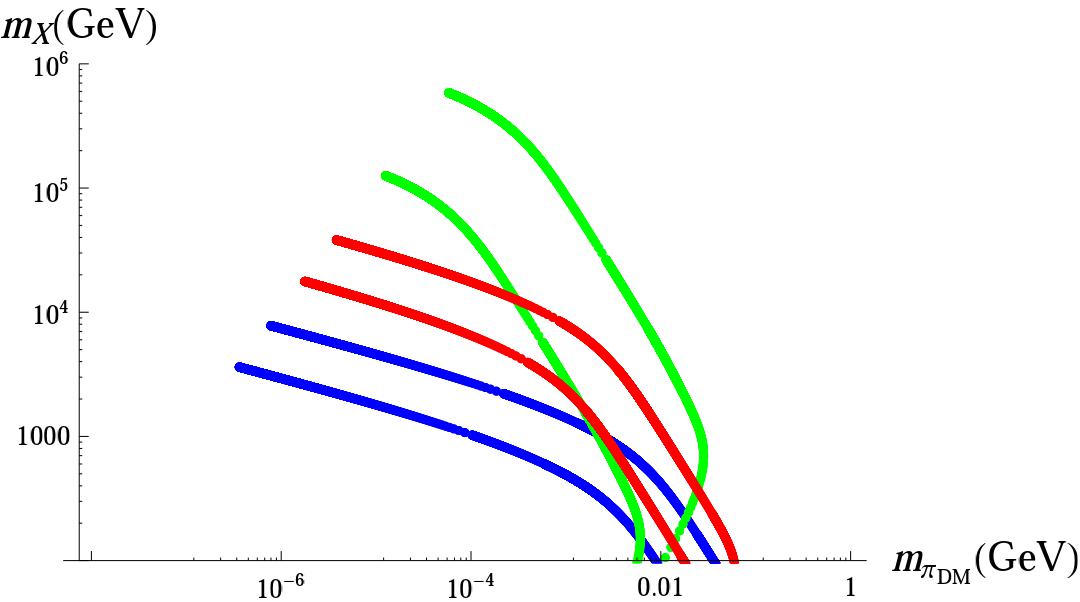}}
\caption{We plot $m_X$ vs $m_{\pi_{DM}}$ 
for the case of $\alpha_{DM}=1$.  We see that all three constraints from clusters, the Milky Way, 
and dwarf galaxies (described in the text) can be met for $m_X$ ranging from a few $100$ GeV to a few TeV since this parameter space falls within all three sets of colored lines.  }
\label{1plot}
\end{center}
\end{figure}

\begin{figure}[h]
\begin{center}
\fbox{\includegraphics[scale=1.4]{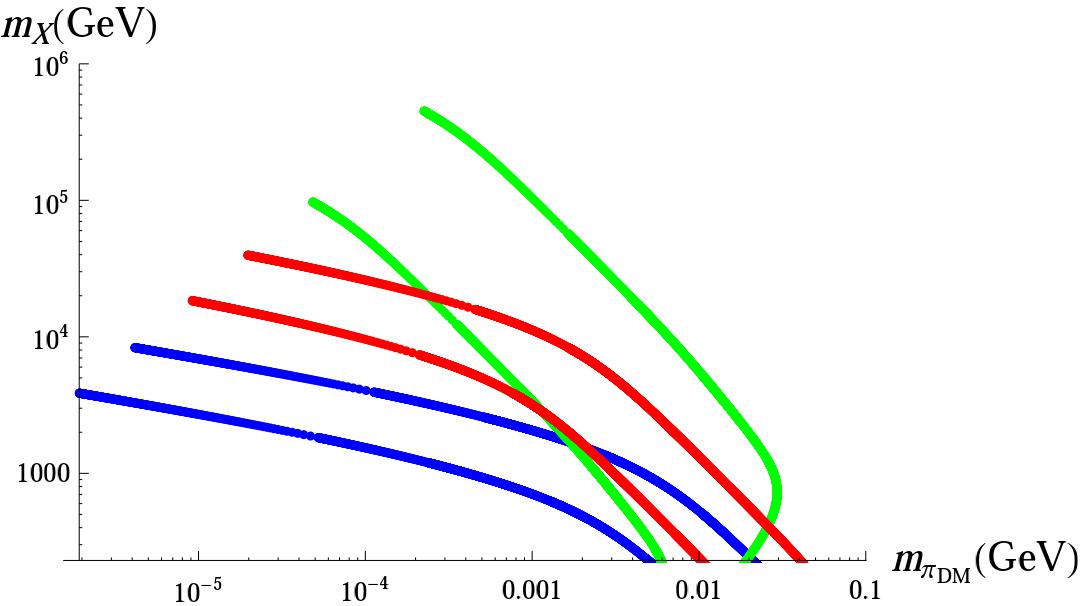}}
\caption{We plot $m_X$ vs $m_{\pi_{DM}}$ 
for the case of $\alpha_{DM}=2$.  We see that all three constraints from clusters, the Milky Way, 
and dwarf galaxies (described in the text) can be met for $m_X$ ranging from a few $100$ GeV to a few TeV since this parameter space falls within all three sets of colored lines.  }
\label{2plot}
\end{center}
\end{figure}

For carrying out the computational method for solving Schr\"{o}dinger's equation exactly as described in \cite{Zurek} with a similar level of computational precision, we make plots of $m_X$ vs $m_{\pi_{DM}}$ for $\alpha_{DM}=1$ and $\alpha_{DM}=2$ via their relationship through the velocity-averaged transfer cross section $<\sigma_T>$ for the interaction described by the potential in Eq. (\ref{potential}).  The plots are given in Figs. (\ref{1plot}) and (\ref{2plot}).

Using the convention of \cite{Zurek}, the plots are described as follows:
\begin{itemize}
\item Green lines going from left to right respectively represent $ \langle \sigma_T \rangle/m_X =10$ and $0.1$ on dwarf scales, required for solving small scale structure anomalies.  
\item Red lines going from left to right respectively represent $\langle \sigma_T \rangle/m_X = 1$ and $0.1$ on Milky Way (MW) scales. 
\item Green lines going from left to right respectively represent $\langle \sigma_T \rangle/m_X = 1$ and $0.1$ on cluster scales. 
\end{itemize}

The above astrophysical upper and lower bounds on  $ \langle \sigma_T \rangle/m_X$, as discussed in \cite{Zurek}, come largely from N-body simulations for a limited number of specific cross sections, so their constraining 
power in our plot 
should not be taken to be extremely stringent.  But the ranges given for $ \langle \sigma_T \rangle/m_X$ are generally what is needed to satisfy observational constraints from structure formation, and we discuss the regions of $m_X$-$m_{\pi_{DM}}$ parameter space that fall within all three ranges (within the bounds of all three sets of colored lines) of $ \langle \sigma_T \rangle/m_X$.   

\bigskip

\begin{center}
{\bf Analysis of Results}
\end{center}

For $\alpha_{DM} \sim 1$, we can see from Figs. (\ref{1plot}) and (\ref{2plot}) that all constraints from cluster, the Milky Way, and dwarf galaxies can be met for $m_X$ ranging from a few $100$ GeV to a 
few TeV, and this range corresponds to $1$ MeV $\lesssim m_{\pi_{DM}} \lesssim 10$ MeV.  For almost all of $m_X < 100$ GeV, we found regions of parameter space that did not grossly disagree with the three 
astrophysical constraints, similar to the results of the resonant regimes in Fig. 6 of \cite{Zurek}.  
We focus our plots on the noteworthy observation that $m_X \gtrsim 10$ TeV does not agree with all three 
constraints in the plots (barring the fact that the tightness of these constraints is open to interpretation).  

Given the numerical results in the previous paragraph, and since $\Lambda_4 \sim m_\chi \leq m_X/4$, one can see that the approximation $m_{0} \sim m_{\pi_{DM}} \ll \Lambda_4$ seems to be a good one, much 
better than the analogous chiral approximation in QCD.  This connection between the constraints on the macroscopic 
astrophysical scale and the microscopic $\pi_{DM}$-$X$ interaction lends support to the viability of the luminogenesis model.  

\begin{figure}
\begin{center}
\fbox{\includegraphics[scale=1.4]{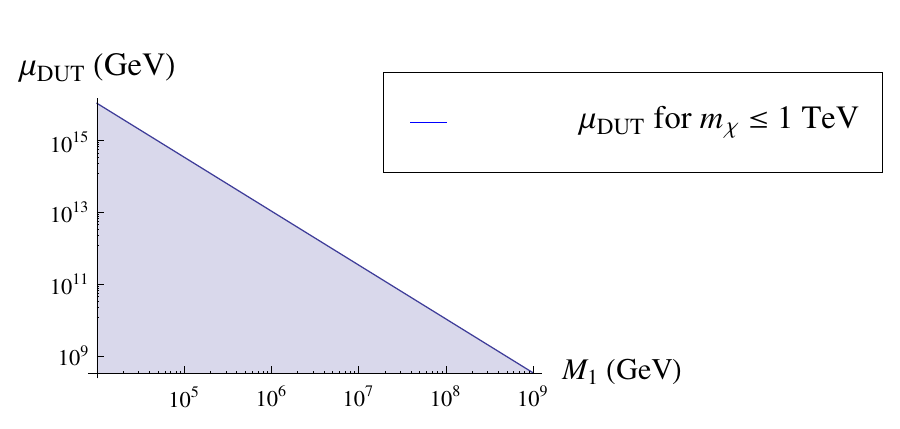}}
\caption{We plot $m_{DUT}$ vs $M_1$ for $m_\chi \leq 1$ TeV.  }
\label{DUTastroDM}
\end{center}
\end{figure}

We now consider the implications of this upper bound on the mass of strongly coupled DM for the luminogenesis model.  Since we saw that $X=(\chi \chi \chi \chi)$ cannot have a mass bigger than a few TeV, and since $m_\chi \leq m_X/4$, we 
see there is an upper bound of $1$ TeV for $m_\chi$.  In Fig. (\ref{DUTastroDM}), we plot $\mu_{DUT}$ vs $M_1$ for this constraint $m_\chi \leq 1$ TeV using Eq. 10 derived from the unification of $SU(2)_L$ and $SU(4)_{DM}$ at the DUT scale in \cite{pqkevin}.  From Fig. (\ref{DUTastroDM}), we see that $\mu_{DUT} \leq 10^{16}$ GeV in order for this astrophysical upper bound 
for $m_\chi$ to be 
satisfied, and most of the viable parameter space (the shaded triangle) 
is for values of $\mu_{DUT}$ much less than $10^{16}$ GeV.  Along with this constraint, we also see that $M_1 \leq 10^9$ GeV.  

Using this upper bound on $\mu_{DUT}$ along with Planck's constraints on the scalar spectral index and amplitude, we can also determine upper bounds on the number 
of e-folds and the parameters of 
the potential for inflation (in our case, the Coleman-Weinberg potential we used in \cite{pqkevin}).  We work out the relationships of these parameters under the constraints 
from Planck in Eq. 21 of \cite{pqkevin}, and one can see that the number of e-folds would need to be less than roughly $95$.  

\bigskip
\begin{center}
{\bf Conclusion}
\end{center}

In this work, we apply astrophysical constraints from clusters, dwarf galaxies, and the Milky Way obtained from N-body simulations for self-interacting dark matter in order to 
solve the structure formation problems known as the cusp-vs-core problem, the missing satellite problem, and the too-big-to-fail problem.  Using the methodology of 
\cite{Zurek}, we assume a Yukawa-type 
interaction between dark matter and a light mediator for a strongly coupled interaction.  In particular, we examine the luminogenesis model and the interaction between 
the bosonic and baryonic CHIMP of $SU(4)$, $X = (\chi \chi \chi \chi)$, and the light mediator $\pi_{DM}$ from chiral symmetry breaking.  

From the astrophysical constraints, for $m_X > 100$ GeV, we saw that $m_X$ can range from a few $100$ GeV to a 
few TeV, and this range corresponds to $1$ MeV $\lesssim m_{\pi_{DM}} \lesssim 10$ MeV.  These values imply a much better chiral approximation $m_0 \ll \Lambda_4$ than 
the analogous approximation in QCD. 
From our analysis, the upper bound of $m_X<$ a 
few TeV led to the upper bounds $\mu_{DUT} \leq 10^{16}$ GeV and $M_1 \leq 10^9$ GeV.  Using the upper bound on $\mu_{DUT}$ along 
with Planck's constraints on inflation, we can obtain upper 
bounds for the parameters of 
an inflation potential and the number of e-folds.  For our Coleman-Weinberg model, the number of e-folds would need to be less than roughly $95$.

With the application of structure formation constraints on the self-interacting dark matter of the luminogenesis model, we have found an interesting observational connection 
between the macroscopic and the microscopic.  We have established a unique connection between observational astrophysics, the nature of dark matter, and inflation, 
and we hope to exploit 
more connections in the future in order to further sharpen our understanding of dark matter and the origins of our universe.

\bigskip
\begin{center}
{\bf Acknowledgements}
\end{center}

PQH was supported in part by the U.S. Department of Energy under
Grant No. DE-FG02-97ER41027 and is supported by the Pirrung Foundation.  
KJL is supported by the Pirrung Foundation.

\bigskip

\end{document}